\makeatother

\documentclass[twocolumn,superscriptaddress,showpacs,aps,prl,floatfix]{revtex4}
\usepackage[latin9]{inputenc}
\usepackage{textcomp}
\usepackage{amsmath}
\usepackage{amssymb}
\usepackage{graphicx}
\usepackage{esint}
\usepackage{bm}
\usepackage{comment}

\newcommand{\bea}{\begin{eqnarray}}
\newcommand{\eea}{\end{eqnarray}}

\def\beq{\begin{equation}}
\def\eeq{\end{equation}}

\begin{document}

\title{Orbital Kondo Spectroscopy in a Double Quantum Dot System}
\author{L. Tosi}
\affiliation{Centro At\'{o}mico Bariloche and Instituto Balseiro, Comisi\'{o}n Nacional
de Energ\'{\i}a At\'{o}mica, 8400 Bariloche, Argentina}
\author{P. Roura-Bas}
\affiliation{Dpto de F\'{\i}sica, Centro At\'{o}mico Constituyentes, Comisi\'{o}n
Nacional de Energ\'{\i}a At\'{o}mica, Buenos Aires, Argentina}
\author{A. A. Aligia}
\affiliation{Centro At\'{o}mico Bariloche and Instituto Balseiro, Comisi\'{o}n Nacional
de Energ\'{\i}a At\'{o}mica, 8400 Bariloche, Argentina}
\date{\today }

\begin{abstract}
We calculate the nonequilibrium conductance of a system of two capacitively coupled
quantum dots, each one connected to its own pair of conducting leads. The system
has been used recently to perform pseudospin spectroscopy by controlling independently
the voltages of the four leads. The pseudospin is defined by the orbital
occupation of one or the other dot. Starting from the SU(4) symmetric point 
of spin and pseudospin degeneracy in the Kondo regime, for an odd number
of electrons in the system, we show how the conductance through each 
dot varies as the symmetry is reduced to SU(2) by a pseudo-Zeeman splitting, and 
as bias voltages are applied to any of the dots.
We analize the expected behavior of the system in general, 
and predict characteristic fingerprint features of the 
SU(4) $\rightarrow$ SU(2) crossover 
that have not been observed so far.

\end{abstract}

\maketitle



The Kondo effect is one of the most studied phenomena in strongly
correlated condensed matter systems \cite{hew} and is still a subject of great interest. 
Originally observed in systems of magnetic impurities in metals, the Kondo
effect has reappeared more recently in the context of semiconductor
quantum-dot (QD) systems, with a single ``impurity'', in which 
an unprecedented control of the parameters could be 
achieved \cite{gold1,cro,gold2,wiel}. 
The effect is
characterized by the emergence of a many-body singlet ground state formed by
the impurity spin and the conduction electrons in the Fermi sea. 
The binding energy of this singlet is of the order of the characteristic Kondo
temperature $T_K$ below which the effects of the ``screening'' of the
impurity spin manifest in different physical properties.

The role of the impurity spin can be replaced by other quantum
degree of freedom (pseudospin) that distinguishes degenerate states, 
such as orbital momentum.  A particularly interesting case is when both a two-fold orbital 
degeneracy and spin degeneracy are present, leading to an 
SU(4) Kondo effect \cite{borda,zar,desint,jari,lim,ander,lipi,buss,fcm,grove,tetta,see,mina,keller}. 
This exotic Kondo effect has been observed in different systems, such as quantum dots in 
carbon nanotubes \cite{jari,lim,ander,lipi,buss,fcm,grove}, silicon nanowires \cite{tetta}, and 
organic molecules deposited on Au(111) \cite{mina}.

\begin{figure}[tbp]
\includegraphics[width=4cm]{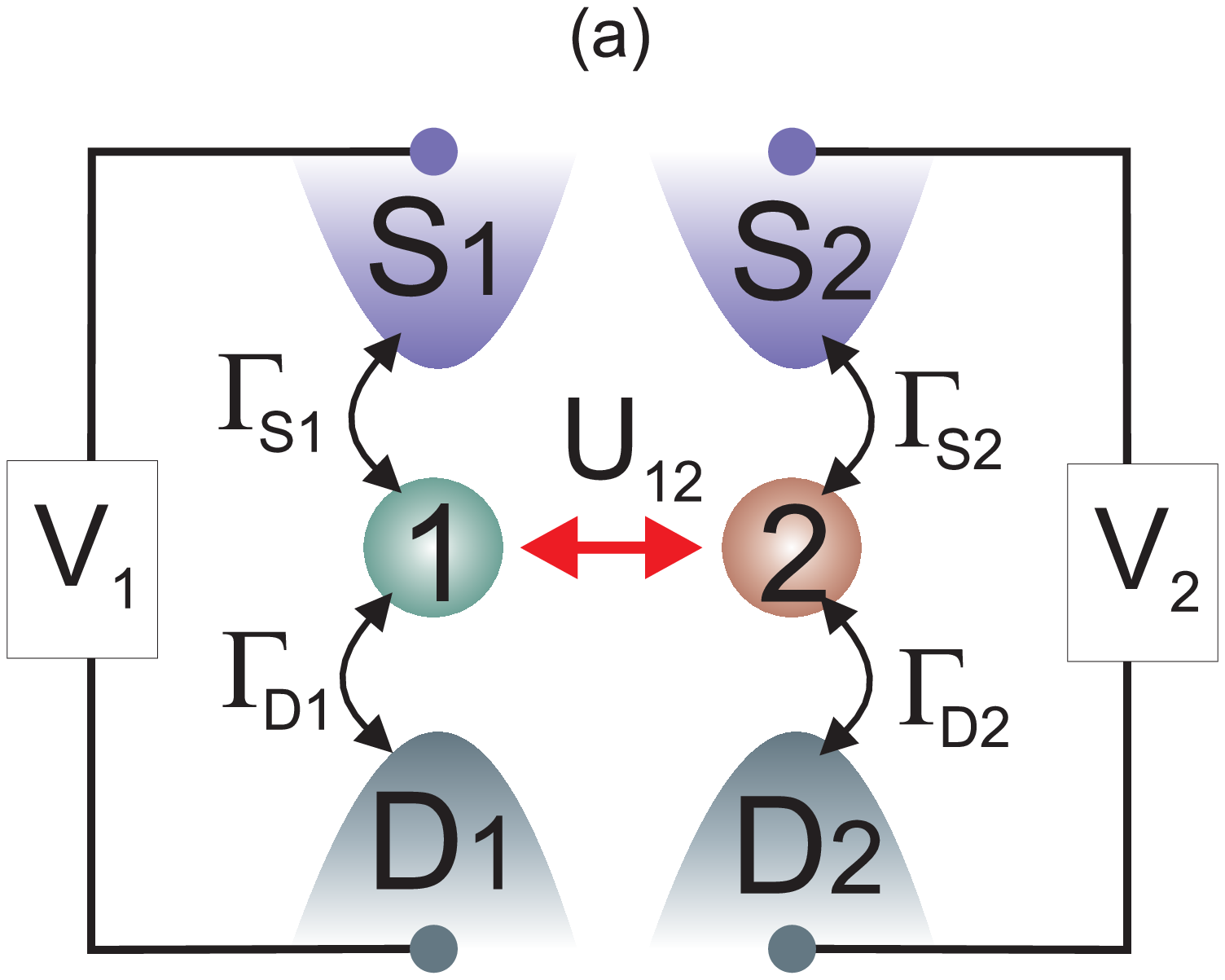}
\includegraphics[width=3.5cm]{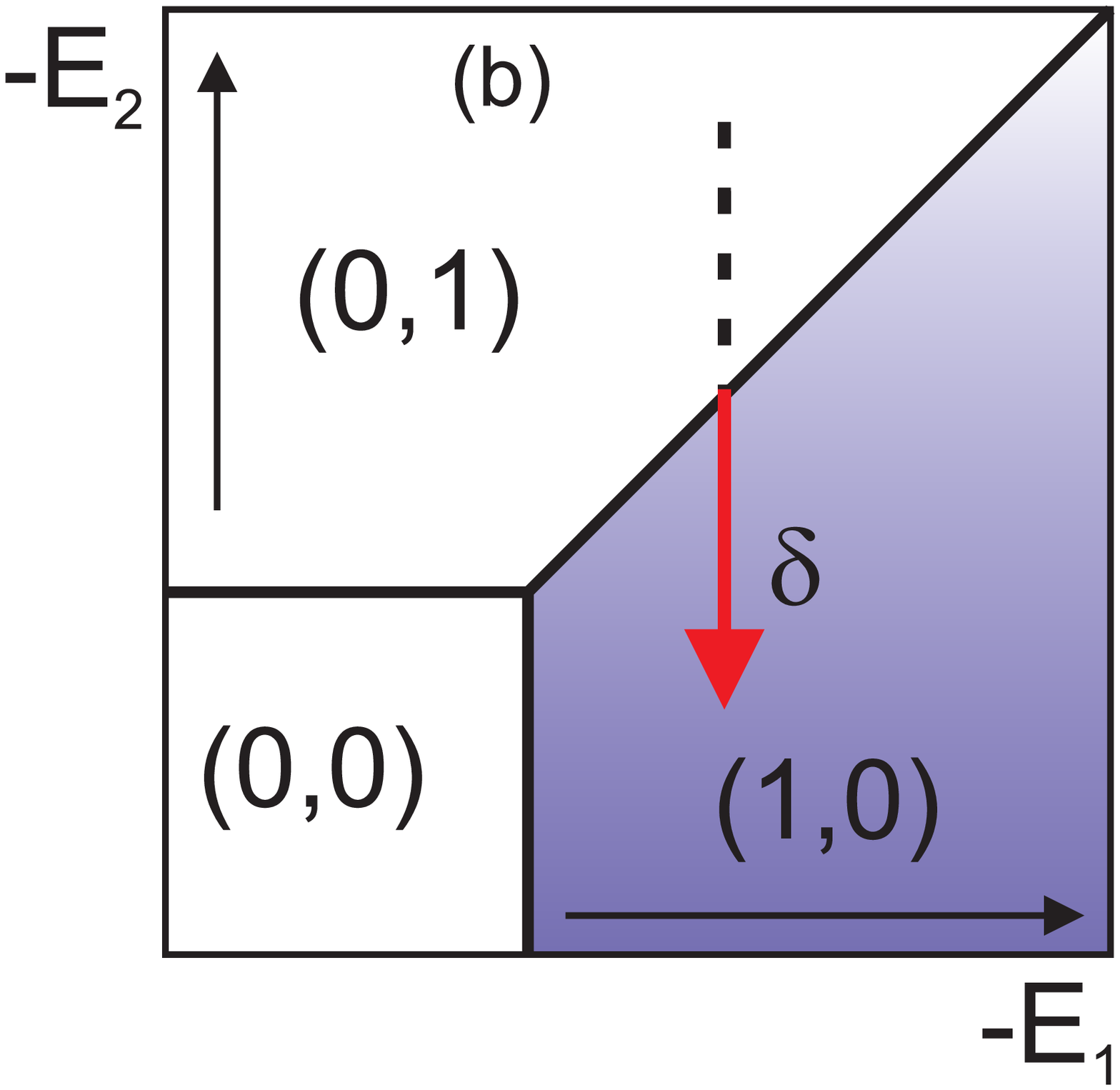}\\
\vspace{0.5cm}
\includegraphics[width=8cm]{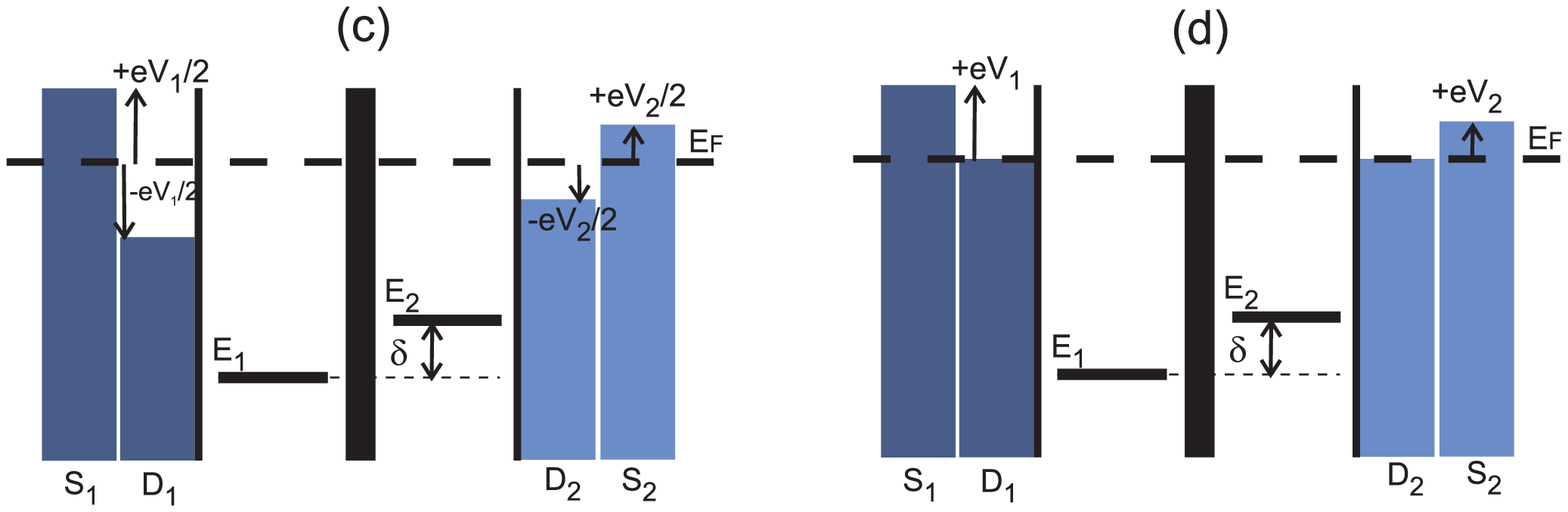}
\caption{(Color online) (a) Scheme of the experimental setup \cite{keller,ama}.
(b) on-site energies of our model and occupations of both quantum dots.
Our convention for application of voltages is in (c) for a symmetric voltage drop and (d)
for voltage applied only to the source leads.}
\label{scheme}
\end{figure}

Recently, a double QD with strong interdot capacitive coupling, and each QD
tunnel-coupled to its own pair of leads has been experimentally \cite{keller,ama} and
theoretically \cite{keller,buss2} studied [see Fig. \ref{scheme} (a)]. The occupation of one QD or the other
plays the role of the pseudospin. These occupations, the tunneling matrix elements
and the voltages at the four leads can be controlled independently. 
While the spin degeneracy can be
broken by a magnetic field, this also affects the conduction leads.
Instead, a pseudo-Zeeman splitting can be applied on the
QDs solely, opening
the exciting possibility to explore in detail the orbital structure of the SU(4) Kondo state 
and how it is changed as the pseudo-Zeeman field reduces the symmetry to SU(2). 

So far, the theoretical study of this system has been concentrated in equilibrium
properties, for which accurate techniques like numerical renormalization group (NRG) and
density-matrix renormalization group can be applied. A much richer physics is expected
in the nonequilibrium situation, which arises for finite bias voltages between the leads connected
to any of the QDs in the experiment, because of the presence of inelastic processes. Unfortunately, 
the theoretical treatment is much more difficult in this case. For one QD, 
the experimental study at finite bias voltages
\cite{grobis} allowed the test of universality and scaling relations within 
different nonequilibrium theories \cite{ogu,scali}.
Here we use the Keldysh formalism within the non-crossing approximation (NCA) \cite{nca,nca2}, which
reproduces well the scaling relations mentioned above \cite{roura} and
was also successfully used to interpret experimental results on a controlled 
crossover between SU(4) and SU(2) Kondo states driven by magnetic field in a nanoscale Si 
transistor \cite{tetta}, and quantum phase transitions involving singlet and 
triplet states \cite{serge}.

Here
we report calculations of
the conductances through both QDs in the general nonequilibrium case.
We describe in particular nontrivial changes in the conductance through one QD
as a voltage is applied to the other. 
We also describe how the spectral densities evolve
under application of different bias voltages.
Fingerprints of the SU(4) $\rightarrow$ SU(2) crossover are predicted.

Our starting model is the SU(4) Anderson model which mixes a singlet configuration with 
two degenerate spin doublets $|i\sigma \rangle$ ($i=1$ or 2) corresponding to one additional electron (or hole)
in QD $i$, through couplings $\Gamma_1=\Gamma_2$ to a continuum of extended states. 
It is described for example in Ref. \onlinecite{see} replacing valley by QD index.    
The symmetry is reduced to SU(2) by a pseudo-Zeeman splitting $\delta=E_2-E_1$,
which raises the energy of a particle in QD2 ($E_2$) with respect to the corresponding one for QD1 ($E_1$). 
The tunnel couplings of each QD to the source and drain leads are $\Gamma_{Si}$ and $\Gamma_{Di}$ 
respectively and we take 
the unit of energy $\Gamma_i=\Gamma_{Si}+\Gamma_{Di}=1$ unless otherwise stated.
$\Gamma_i$ correspond to the total width at half maximum of the spectral density
in the non-interacting system. It is of the order of 20 $\mu$eV in the experiments \cite{keller,ama}
Since charge configurations with two particles are excluded, the model assumes infinite on-site repulsions
$U_i$ and interdot repulsion $U_{12}$. This assumption is not essential in the Kondo regime for one
particle (electron or hole) in the system, which is the focus of our study.
A scheme of the setup and basic parameters is represented in Fig. \ref{scheme}.
In real systems, $\Gamma_1 \ne \Gamma_2$ and SU(4) symmetry is lost even for $\delta=0$.
However, we find that tuning appropriately $\delta$, the equilibrium spectral densities
for both dots $\rho_i(\omega)$ can be made to coincide at low temperatures. This indicates 
that the SU(4) symmetry is recovered as an emergent (approximate) symmetry \cite{eme} at low temperatures.


\begin{figure}[h!]
\includegraphics[width=7cm]{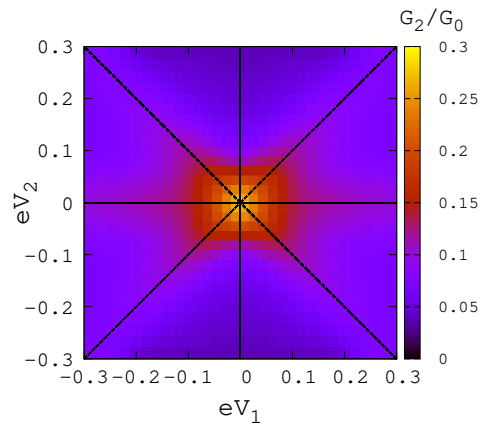}
\caption{(Color online) Conductance of QD2 as a function of $V_1$ and $V_2$ for $\delta=0$, 
$E_1=-4$ and $T=0.005$.}
\label{map0}
\end{figure}

We start reporting the  differential conductances $G_i=dI_i/dV_i$ in the SU(4) symmetric case.
In Fig. \ref{map0} we show $G_2$ as a function of both
$V_i$ for symmetric voltage drops [$V_{Si}=-V_{Di}=V_i/2$, see Fig. \ref{scheme} (c)] 
and coupling to the leads ($\Gamma_{Si}=\Gamma_{Di}$).  
By SU(4) symmetry, $G_1$ 
has the same form interchanging
$V_1$ and $V_2$. For $V_1=V_2=0$, there is a maximum of the conductance due to the 
SU(4) Kondo effect \cite{lim,ander,see}. At temperature $T=0$, this maximum is slightly below $G_0/2$, where
$G_0= 2 e^2/h$, due to some degree of intermediate valence, according to Friedel sum rule \cite{fcm,see} 
(the filling is slightly below 1/4). 
Application of either $V_1$ or $V_2$ tends to destroy the Kondo effect
and the conductance decreases. Note that application of $V_2$ has a stronger effect on decreasing $G_2$ than $V_1$.
In fact for $V_2=0$ and any $V_1$, one expects that the spin Kondo effect on QD2 still remains, 
although weakened, and this is consistent with our results.
As in the usual SU(2) Kondo effect, $G_i (V_i)$ drops to $G_i (0)/2$ at a bias voltage such that
$eV_i \approx T_K^{SU(4)}$, where $T_K^{SU(4)}$ is the Kondo temperature for $\delta=0$ as
discussed below.
For our parameters, $T_K^{SU(4)} \approx 0.02$ and it increases to near 0.3 if $E_1$ is changed from -4 to -2.
Since experimentally temperatures $T \approx 0.1$ can be reached, and $E_1$ can be tuned, a wide range of ratios 
$T/T_K^{SU(4)}$ is accessible. 

Non-trivial correlation effects between both QDs are apparent in the fact that $G_i$ increases on the lines
$V_1 = \pm V_2$. This is related to the evolution of the spectral densities $\rho_i(\omega)$ as both $V_i$ 
are varied. We find that keeping $V_2=0$ and increasing $V_1$ (or conversely) the Kondo peak at $\omega=0$ 
in $\rho_i(\omega)$ is weakened and two peaks at $\omega \approx \pm eV_1/2$ split from it. In the general 
case, when both $V_i \ne 0$, four peaks are present in both $\rho_i(\omega)$ for $\omega \approx \pm eV_i/2$.
When $V_1= \pm V_2$ these peaks merge in two more intense peaks and therefore an increase in both $G_i$
is expected.

\begin{figure}[h!]
\includegraphics[width=8cm]{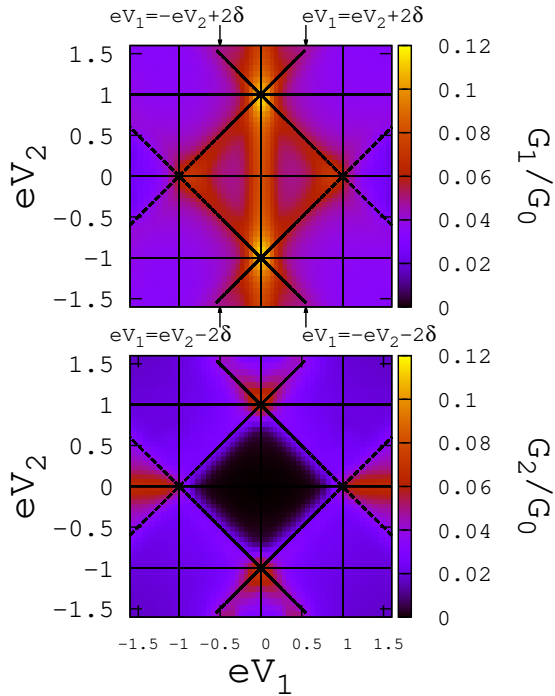}
\caption{(Color online) Conductance of (a) QD1 and (b) QD2 as a function of $V_1$ and $V_2$ for 
$\delta=E_2-E_1=0.5$, $E_1=-4$ and $T=0.005$.}
\label{map2}
\end{figure}

\begin{figure}[h!]
\includegraphics[width=4cm]{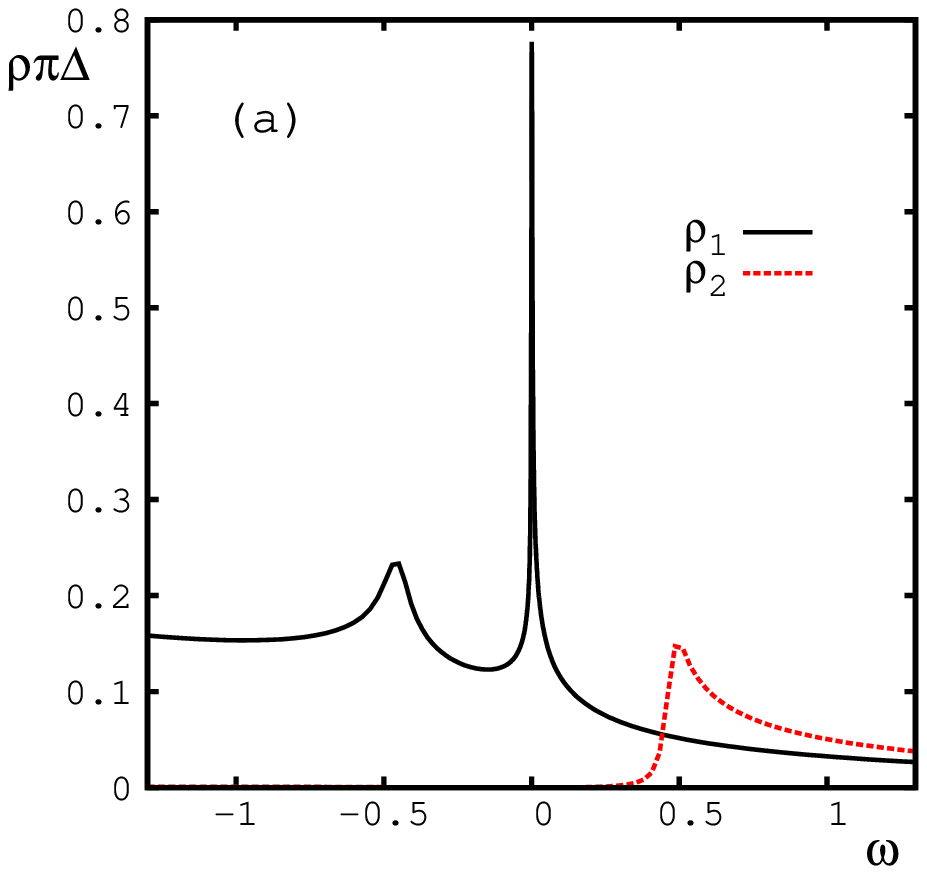}
\includegraphics[width=4.3cm]{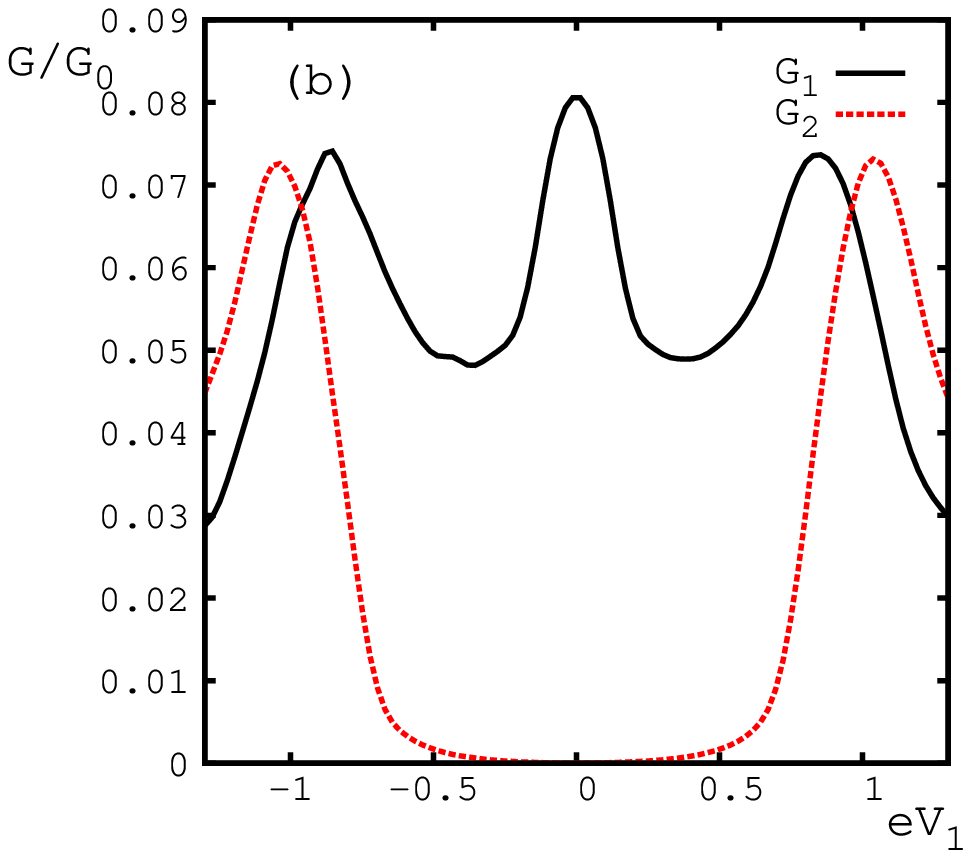}
\caption{(Color online) (a) spectral densities for QD1 (black solid line)
and QD2 (red dashed line). (b) $G_i$ as a 
function its bias voltage $V_i$ keeping the other bias voltage 0.
Parameters as in Fig. \ref{map2}. $\Delta=\Gamma/2$}
\label{rho}
\end{figure}

In Fig. \ref{map2} we show how the $G_i$ change when a finite pseudo-Zeeman splitting $\delta$ 
is introduced. It is known that the spectral density of the dot with lower energy $\rho_1(\omega)$
has still the Kondo peak near $\omega=0$ and an additional peak for $\omega \approx -\delta$, while 
$\rho_2(\omega)$ has only a peak for $\omega \approx \delta$ (see Fig. \ref{rho}) \cite{fcm,lim}. 
As a consequence,
only $G_1$ has a peak near $V_1=V_2=0$, while $G_2$ is vanishingly small at that point. The energy
scale of the variation of $G_1$ with $V_1$ is again given by the Kondo temperature $T_K$, but it is 
smaller than that of the SU(4) case. We have found that the binding energy of the singlet ground state obtained
from a simple variational calculation can be 
described by the following expression \cite{fcm}

\begin{eqnarray}
\frac{T_K(\delta)}{T_K(0)}=\sqrt{1+\tilde{\delta}/d +\tilde{\delta}^2}-\tilde{\delta}, \nonumber \\
\tilde{\delta}=\frac{\delta}{2 T_K(0)} {\rm ,~   } d=\frac{D}{2 T_K(0)}
\label{tk},
\end{eqnarray}
where $T_K(0)=T_K^{SU(4)} \approx D {\rm exp}[\pi E_1/(2 \Gamma)]$ 
and $D$ is half the band width 
(we took $D=10$). The width of the Kondo peak calculated within the
NCA agrees remarkably well with this expression \cite{fcm}.
Specifically for $E_1=-4$ the total width of the Kondo peak
in the spectral density is found to be $1.2 T_K(\delta)$.  

We find that for $T<T_{K}({\delta )}$ and $V_{1}>T_{K}({\delta )}$, 
$G_{1}(V_{1})$ for $V_{2}=0$ presents a structure with three peaks 
(see Fig. \ref{rho} (b)) which has not been observed experimentally yet and is
characteristic of the SU(4) $\rightarrow$ SU(2) crossover \cite{note}. 
Since $T_{K} (\delta)$
varies over several orders of magnitude, we believe that an experimental
observation of this fingerprint of the crossover is near the present
experimental possibilities \cite{keller,ama}. It would also be interesting 
to test the dependence of $T_{K}$ on $\delta$. 

Another apparent feature is the increase of both
conductances along the lines $eV_{1}=\pm eV_{2}\pm 2\delta$ (shown dashed 
in Fig. \ref{map2}). This can be
understood from the onset of cotunneling events near this equalities \cite{will}. 
Let us assume first that $V_{1}=V_{2}=0$. The cotunneling event in which the electron that occupies
QD1 jumps to its source (S1) or drain (D1) lead and an electron from S2 or
D2 jumps to QD2 is inhibited because of the energy cost $\delta $. However,
when $e|V_{2}|$ reaches $2\delta $, an event of this type becomes possible,
in which as a net result an electron flows from S2 to D2 or conversely
depending on the sign of $V_{2}$, and another electron moves from QD1 to QD2
with a possible spin flip. In a second event the electron of QD2 jumps to
its lead of less energy and an electron from S1 or D1 jumps to QD1, leaving
the QDs in the same charge configuration as initially. This results in an increase
of the current flow $I_{2}$ and thus to a peak in the conductance $G_{2}$. On average $I_{1}=0$.
However, a small $|V_{1}|$ breaks the symmetry between S1 and D1 in the
above events, leading to a large $G_{1}$ also. A similar reasoning can be
followed for non zero $V_{1}$.

More insight into the structure of the nonequilibrium conductance is
obtained from the spectral densities  $\rho _{i}(\omega )$. 
At equilibrium and low temperatures, both $\rho _{i}(\omega )$ have
a Kondo peak slightly above the Fermi energy (which we take as the origin of energies)
for $\delta < T_{K}^{SU(4)}$ \cite{fcm}, while for $\delta > T_{K}^{SU(4)}$, 
as seen in Fig. \ref{rho} (a), the Kondo peak in $\rho _{1}(\omega )$ moves to  
the Fermi energy and an inelastic peak near $-\delta $
appears ($\delta=0.5$ in the figure). The width of the Kondo peak is $\sim T_{K}({\delta )}$.
Instead  
$\rho_{2}(\omega )$ has only an inelastic peak near energy $\delta $. We find
that the width of both inelastic peaks is of the order of  $T_{K}^{SU(4)}$
for small $\delta $ (but $\delta > T_{K}^{SU(4)}$ in order to ensure that the inelastic
peak is split from the Kondo peak) and increases with increasing  $\delta$.
This behavior is reminiscent of the evolution of the peaks of the ordinary
SU(2) Kondo model under an applied magnetic field, which has been studied by
Bethe ansatz techniques \cite{moore}. 

\begin{figure}[h!]
\includegraphics[clip, width=5cm]{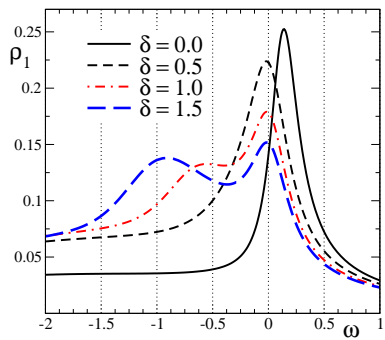}
\caption{(Color online) spectral density for QD1 for $E_1=-2.5$, $T=0.1$ 
and several values of $\delta$.}
\label{rho2}
\end{figure}

The equilibrium spectral densities can be investigated by orbital spectroscopy
controlling the parameters so that the configuration is similar to that used in 
scanning tunneling spectroscopy (STS) . Specifically if $\Gamma _{Si} \gg \Gamma _{Di}$,
and only the potential at one of the drains $V_{Di}$ is displaced from the Fermi level, 
then the dots are in equilibrium
with the source leads and for $T \ll T_{K}({\delta )}$,
$G_i \propto \rho _{i}(eV_i)$. Our calculations show that a ratio 
$\Gamma _{Si} / \Gamma _{Di}=9$ is enough to reach this STS regime. This property 
was used to study the equilibrium spectral density
and to compare it with that resulting from an NRG calculation
for a case in which both spin and pseudospin Zeeman terms were present \cite{keller}.
However, this destroys the Kondo effect and the two-peak structure like that shown in 
Fig. \ref{rho} (a) remains unexplored. 
In Fig. \ref{rho2} we show the evolution of the spectral density starting from the 
SU(4) case and increasing $\delta$ for parameters reached experimentally in recent work 
\cite{keller,ama}, in particular $\Gamma_i=20 \mu$eV, $T=23$ mK. While the peaks become sharper 
at lower temperature, the displacement of the Kondo peak to the Fermi energy from above,
and the splitting of the inelastic peak as $\delta$ increases can be clearly seen.
Due to the limitations of resolution of NRG at 
finite energies \cite{vau,freyn} our NCA results are a useful complement at equilibrium \cite{note2} 
and have the advantage that they can be extended to the nonequilibrium situation.

In general and  particularly for a symmetric voltage drop
the spectral densities change under application of
bias voltages $V_{i}$. Assuming as a first crude
approximation that the $\rho _{i}(\omega)$ are fixed, one expect that
$G_{1}(V_{1})$ has a peak at $V_{1}=0$ corresponding to the Kondo peak in  
$\rho _{1}(\omega )$, and two peaks at $V_{1}=\pm 2\delta /e$ corresponding
to the inelastic peak of $\rho _{1}(\omega )$. This is in fact what happens
for $V_{2}=0$ [see Fig. \ref{rho} (b)] but not for $V_{2}\neq 0$ [see
Fig. \ref{map2} (a)]. Similarly one expects only inelastic peaks at  
$V_{2}=\pm 2\delta /e$ for $G_{2}(V_{2})$, as it happens for $V_{1}=0$ but
not for $V_{1}\neq 0$. 

\begin{figure}[h!]
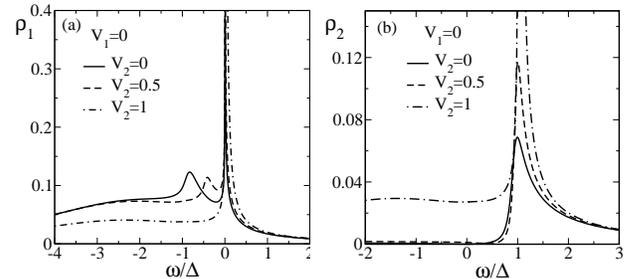

\includegraphics[width=4.0cm]{rho-1-vs-v2.eps}
\includegraphics[width=4.0cm]{rho-2-vs-v2.eps} 
\caption{
Spectral density of (a) QD1  and (b) QD2  as a function of
frequency for $E_1=-3$, $\delta=1$ and several bias voltages at source 2.}
\label{rhonone}
\end{figure}

The differences with the expected behavior for rigid bands when both $V_i \ne 0$
are due to changes in the spectral weight with respect to the equilibrium case.
To illustrate these changes we consider the nonequilibrium situation
represented in Fig. 4 (d) of Ref. \onlinecite{ama}, in which the coupling to the 
source leads is larger and the voltages are applied only in one of these sources $V_{Si}$,
keeping the other three voltages at zero. Specifically we keep  
$\Gamma_{i}=1$ but use $\Gamma _{S1} / \Gamma _{D1}=3$ and $\Gamma _{S2} / \Gamma _{D2}=12$,
as described in the supplementary material of Ref. \onlinecite{ama}. We also changed
$E_1=-3$ and $\delta=1$ to correspond approximately to the experimental parameters.
The evolution of $\rho _{1}(\omega)$ with $V_{S2}$ is shown in Fig. \ref{rhonone} (a).
At equilibrium ($V_{S2}=0$), the spectral density of QD1 has the two peaks mentioned above. The
inelastic peak can be understood as a mixture of the ground state for zero hopping with an excited 
state in which the electron at QD1 is displaced to QD2 and an an electron from S2 is displaced to
D1. Both states are connected in second order in the lead-QDs hopping. The excitation energy is 
$\delta$. As a consequence of this mixture, when an electron is destroyed in QD1, there is a finite 
probability of leaving an excited state with energy $\delta$. This leads to a peak at 
$-\delta$ in $\rho _{1}(\omega)$. When the chemical potential 
at S2 is increased, the excitation energy decreases and the peak displaces towards the Fermi energy.
When this potential reaches $\delta$, the inelastic peak merges with the elastic one and this leads 
to a peak in $G_{1}(V_{1})$ at $V_{1}=0$, even at temperatures above $T_{K}({\delta )}$ for
which the original elastic peak disappears. This agrees with the result presented 
in Fig. 4 (d) of Ref. \onlinecite{ama}. We obtain a qualitative agreement with experiment, 
but the ratio of intensities is larger in our case. This might be due uncertainties in the ratio
$E_1 / \Gamma _i$ or to fluctuations in $\delta$ introduced by decoherence effects \cite{pri}.

A similar reasoning as above can be followed for a symmetric voltage drop and brings an alternative 
explanation of the increase in intensity along the lines $eV_{1}=\pm eV_{2}\pm 2\delta$ 
displayed in Fig. \ref{map2}.

In Fig. \ref{rhonone} (b) we show how $\rho _{2}(\omega)$ changes with $V_{S2}$. In contrast to 
$\rho _{1}(\omega)$, much of the spectral weight lies above the Fermi energy. Therefore its 
magnitude is proportional to the amount of the singlet configuration without particles in the ground
state, or in other words, to the degree of intermediate valence. We observe that 
$\rho _{2}(\omega)$ increases as $V_{S2}$ approaches $\delta$.


In summary, we predict the values of 
the conductance through any of two capacitively coupled QDs as the voltage through
any of them is varied. We believe that our results are important to stimulate
further experimental research along the lines of recent pseudospin-resolved
transport measurements \cite{keller,ama}. In particular, the presence of 
three peaks in $G_{1}(V_{1})$ for $V_2=0$ (or two peaks in an STS configuration)
with one of them at $(V_{1})=0$ is characteristic of the SU(4) $\rightarrow$ SU(2) 
crossover \cite{note}. This has not been observed in experiment yet \cite{keller,ama}.
However, giving the large experimental possibilities of tuning the parameters and the
particular sensitivity of $T_{K}({\delta)}$ with the pseudo Zeeman splitting
$\delta$ we believe that it can be observed in the near future. An experimental study
of the dependence of the lowest energy scale $T_{K}$ with the pseudo-Zeeman splitting 
$\delta$ would also contribute to our present understanding of the SU(4) $\rightarrow$ SU(2) 
crossover.


We thank A. J. Keller, G. Zar\'and, S. Amasha, D. Goldhaber-Gordon and A. M. Llois  
for useful discussions. 
This work was partially supported by 
PIP 112-200801-01821 and PIP-00273 of CONICET, and PICT 10-1060 and PICT 2011-1187 of the
ANPCyT, Argentina.


\begin{thebibliography}{99}
\bibitem{note} A differential conductance with three peaks has been also observed
in transport through $C_{60}$ quantum dots in which triplet states are important 
\cite{roch,serge} and explained by NCA calculations \cite{serge,st1,st2}. However,
the dependence of gate voltage of the position of the peaks is quite 
different \cite{serge}.
\bibitem{note2}  An example of the limitations of NRG to detect finite-energy features
is the plateau at
intermediate temperatures observed in transport through C$_{60}$ molecules
for gate voltages for which triplet states are important \cite{roch,serge},
which was missed in early NRG studies, but captured by the NCA \cite{st1,st2}.
More recent NRG\ calculations using tricks to improve the resolution \cite{freyn}, 
have confirmed this plateau \cite{serge}.
\end{thebibliography}
\end{document}